\documentclass[journal,draftclsnofoot,onecolumn,12pt]{IEEEtran}
\ifCLASSINFOpdf
\else
\fi
\usepackage[cmex10]{amsmath}
\usepackage{array}
\usepackage{mdwmath}
\usepackage{mdwtab}
\usepackage{fixltx2e}
\usepackage{graphicx}
\usepackage{amssymb}
\usepackage{amsbsy} 
\usepackage{float}
\usepackage{balance}
\usepackage{subfigure}
\usepackage{url}
\newfloat{longequation}{b}{ext}

\newtheorem{thm}{Theorem}[section]

\newtheorem{lem}{Lemma}[section]

\newtheorem{rem}{Remark}[section]

\begin{document}

\title{Interference Mitigation Techniques \\for Clustered Multicell Joint Decoding Systems}

\author{Symeon~Chatzinotas,
        Bj\"{o}rn~Ottersten
        % <-this % stops a space
\thanks{S. Chatzinotas and B. Ottersten are with the Interdisciplinary Centre for Security, Reliability and Trust (SnT), University of Luxembourg  (http://www.securityandtrust.lu). B. Ottersten is also with the Royal Institute of Technology (KTH), Sweden.
 e-mail: \textbraceleft Symeon.Chatzinotas, Bjorn.Ottersten\textbraceright@uni.lu.
 Part of this work will be presented in IEEE\ WCNC 2011.
 }% <-this % stops a space
%\thanks{Manuscript received April 19, 2007; revised January 11, 2008.}
}

\maketitle

\begin{abstract}
%\boldmath
Multicell joint processing has originated from information-theoretic principles as a means of reaching the fundamental capacity limits of cellular networks. However, global multicell joint decoding is highly complex and in practice clusters of cooperating Base Stations constitute a more realistic scenario. In this direction, the mitigation of intercluster interference rises as a critical factor towards achieving the promised throughput gains. In this paper, two intercluster interference mitigation techniques are investigated and compared, namely interference alignment and resource division multiple access. The cases of global multicell joint processing and cochannel interference allowance are also considered as an upper and lower bound to the interference alignment scheme respectively. Each case is modelled and analyzed using the per-cell ergodic sum-rate throughput as a figure of merit. In this process, the asymptotic eigenvalue distribution of the channel covariance matrices is analytically derived based on free-probabilistic arguments in order to quantify the sum-rate throughput.  Using numerical results, it is established that resource division multiple access is preferable for dense cellular systems, while cochannel interference allowance is advantageous for highly sparse cellular systems. Interference alignment provides superior performance for average to sparse cellular systems on the expense of higher complexity. \end{abstract}


% \fi
\IEEEpeerreviewmaketitle

\section{Introduction}
\IEEEPARstart{C}{urrently} cellular networks carry the main bulk of wireless traffic and as a result they risk being saturated considering the ever increasing traffic imposed by internet data services. In this context, the academic community in collaboration with industry and standardization bodies have been investigating innovative network architectures and communication techniques which   can overcome the interference-limited nature of cellular systems. The paradigm of multicell joint processing has risen as a promising way of overcoming those limitations and has since gained increasing momentum which lead from theoretical research to testbed implementations \cite{EASYC}. Furthermore, the recent inclusion of CoMP (Coordinated Multiple Point) techniques in LTE-Advanced \cite{3GPP2010}   serves as a reinforcement of the latter statement. 

Multicell joint processing is based on the idea that signal processing does not take place at individual Base Stations (BSs), but at a central processor which can jointly serve the User Terminals (UTs) of multiple cells through the spatially distributed BSs. It should be noted that the main concept of multicell joint processing is closely connected to the rationale behind Network MIMO\ and Distributed Antenna Systems (DAS) and those three terms are often utilized interchangeably in the literature.  According to the global multicell joint processing, all the BSs of a large cellular system are assumed to be interconnected to a single central processor through an extended backhaul. However, the computational requirements of such a processor and the large investment needed for backhaul links have hindered its realization. On the other hand, clustered multicell joint processing utilizes multiple signal processors in order to form BS clusters of limited size, but this localized cooperation introduces intercluster interference into the system, which has to be mitigated in order to harvest the full potential of multicell joint processing. In this direction, reuse of time or frequency channel resources (resource division multiple access) could provide the necessary spatial separation amongst clusters, an approach which basically mimics the principles of the traditional cellular paradigm only on a cluster scale. Another alternative would be to simply tolerate intercluster signals as cochannel interference, but obviously this scheme becomes problematic in highly dense systems. 

Taking all this into account, the current paper considers the uplink of a clustered Multicell Joint Decoding (MJD) system and proposes a new communication strategy for mitigating intercluster interference using Interference Alignment (IA). More specifically, the main contributions herein are:
\begin{enumerate}
        \item the channel modelling of a clustered MJD system with IA as intercluster interference mitigation technique,
        \item the analytical derivation of the ergodic throughput based on free probabilistic arguments in the R-transform domain,
        \item the analytical comparison with the upper bound of global MJD, the Resource Division Multiple Access (RDMA) scheme and the lower bound of clustered MJD with Cochannel Interference allowance (CI),
        \item the comparison of the derived closed-form expressions with Monte Carlo simulations and the performance evaluation using numerical results.
\end{enumerate}

The remainder of this paper is structured as follows: Section \ref{sec: related work} reviews in detail prior work in the areas of clustered MJD and IA. Section \ref{sec: eigenvalue distribution analysis} describes the channel modelling, free probability derivations and throughput results for the following cases: a) global MJD, b) IA, c) RDMA and d) CI. Section \ref{sec: numerical results} displays the accuracy of the analysis by comparing to Monte Carlo simulations and evaluates the effect of various system parameters in the throughput performance of clustered MJD. Section \ref{sec: conclusion} concludes the paper.

\subsection{Notation}
Throughout the formulations of this paper, \(\mathbb{E}[\cdot]\) denotes  expectation,  \(\left(\cdot\right)^H\) denotes the conjugate matrix transpose, \(\left(\cdot\right)^T\) denotes the matrix transpose, \(\odot\) denotes the Hadamard product and \(\otimes\) denotes the Kronecker product. The Frobenius norm of a matrix or vector is denoted by \(\left\Vert \cdot\right\Vert\) and the delta function by 
$\delta(\cdot)$. $\mathbf{I}_n$ denotes an $n\times n$ identity matrix, $\mathbb{I}_{n\times m}$ an $n\times m$ matrix of ones, $\mathbf{0}$ a zero matrix and $\mathbf{G}_{n\times m}\sim\mathcal{CN}(\mathbf 0,\mathbf I_n) $ denotes $n\times m$ Gaussian matrix with entries drawn form a $\mathcal{CN}(0,1) $ distribution. 
% $[\mathbf{X}_1\ \mathbf{X}_2\ldots\mathbf{X}_n]$ symbolizes matrix or vector concatenation. 
The figure of merit analyzed and compared throughout this paper is the ergodic per-cell sum-rate throughput\footnote{The term throughput is used instead of capacity since the described techniques lead to achievable sum-rates except for MJD which leads to MIMO MAC capacity.}.

\section{Related Work}
\label{sec: related work}

\subsection{ Multicell Joint Decoding}
This section reviews the literature on MJD systems by describing the evolution of global MJD models and subsequently focusing on clustered MJD approaches.
\subsubsection{Global MJD}
It was almost three decades ago when the paradigm of global MJD was initially proposed  in two seminal papers \cite{Hanly1993,Wyner1994},  promising large capacity enhancements. The main idea behind global MJD is the existence of a central processor (a.k.a. ``hyper-receiver'') which is interconnected to all the BSs through a backhaul of wideband, delayless and error-free links. The central processor is assumed to have perfect Channel State Information (CSI) about all the wireless links of the system. The optimal communication strategy is superposition coding at the UTs and successive interference cancellation at the central processor. As a result, the central processor is able to jointly decode all the UTs of the system, rendering the concept of intercell interference void. 

Since then, the initial results were extended and modified by the research community for more practical propagation environments, transmission techniques and backhaul infrastructures in an attempt to more accurately quantify the performance gain. More specifically, it was demonstrated in \cite{Somekh2000} that Rayleigh fading promotes multiuser diversity which is beneficial for the ergodic capacity performance. Subsequently, realistic path-loss models and user distributions were investigated in \cite{Chatzinotas_letter,Chatzinotas_SPAWC} providing closed-form ergodic capacity expressions based on the cell size, path loss exponent and geographical distribution of UTs. The beneficial effect of MIMO links was established in \cite{Aktas2006,Chatzinotas_ISWCS}, where a linear scaling of the ergodic per-cell sum-rate capacity with the number of BS antennas was shown. However, correlation between multiple antennas has an adverse effect as shown in \cite{Chatzinotas_JWCOM}, especially when correlation affects the BS-side. Imperfect backhaul connectivity has also a negative effect on the capacity performance as quantified in \cite{Simeone2008}. MJD has been also considered in combination with DS-CDMA \cite{Somekh2007a}, where chips act as multiple dimensions. Finally, linear MMSE filtering \cite{Ng2007,Chatzinotas_VTC10} followed by single-user decoding has been considered as an alternative to the optimal multiuser decoder which requires computationally-complex successive interference cancellation. 
\subsubsection{Clustered MJD}
Clustered MJD  is based on forming groups of $M$ adjacent BSs (clusters) interconnected to a cluster processor. As a result, it can be seen as an intermediate state between traditional cellular systems (\(M=1\)) and global MJD (\(M=\infty\)). The advantage of clustered MJD lies on the fact that both the size of the backhaul network and the number of UTs to be jointly processed decrease. The benefit is twofold; firstly, the extent of the backhaul network is reduced and secondly, the computational requirements of MJD (which depend on the number of UTs) are lower. The disadvantage is that the sum-rate capacity performance is degraded by intercluster interference, especially affecting the individual rates of cluster-edge UTs. This impairment can be tackled using a number of techniques as described here. The simplest approach is to just treat it as cochannel interference and evaluate its effect on the system capacity as in \cite{Chatzinotas_JCOM2}. An alternative would be to use RDMA, namely to split the time or frequency resources into orthogonal parts dedicated to cluster-edge cells \cite{Katranaras2009}. This approach eliminates intercluster interference but at the same time limits the available degrees of freedom. In DS-CDMA MJD systems, knowledge of the interfering codebooks has been also used to mitigate intercluster interference \cite{Somekh2007a}. Finally, antenna selection schemes were investigated as a simple way of reducing the number of intercluster interferers \cite{Choi2007}.

\subsection{Interference Alignment}
This section reviews the basic principles  of IA and subsequently describes existing applications of IA on cellular networks. 
\subsubsection{IA Preliminaries}
IA  has been shown to achieve the degrees of freedom (dofs) for a range of interference channels \cite{Cadambe2008,Cadambe2009,Jafar2008}. Its principle is based on aligning the interference on a signal subspace with respect to the non-intended receiver, so that it can be easily filtered out by sacrificing some signal dimensions. The advantage is that this alignment does not affect the randomness of the signals and the available dimensions with respect to the intended receiver. The disadvantage is that the filtering at the non-intended receiver removes the signal energy in the interference subspace and reduces the achievable rate. The fundamental assumptions which render IA feasible are that there are multiple available dimensions (space, frequency, time or code) and that the transmitter is aware of the CSI towards the non-intended receiver. The exact number of needed dimensions and the precoding vectors to achieve IA are rather cumbersome to compute, but a number of approaches have been presented in the literature towards this end \cite{Gomadam2008,Yetis2010,Tresch2009}. 
\subsubsection{IA and Cellular Networks}
IA has been also investigated in the context of cellular networks, showing that it can effectively suppress cochannel interference \cite{Suh2008,Tresch2009}. More specifically, the downlink of an OFDMA cellular network with clustered BS cooperation is considered in \cite{Tresch2009b}, where IA is employed to suppress intracluster interference while intercluster interference has to be tolerated as noise. Using simulations, it is shown therein that even with unit multiplexing gain the throughput performance is increased compared to a frequency reuse scheme, especially for the cluster-centre UTs. In a similar setting, the authors in \cite{Da2011} propose an IA-based resource allocation scheme which jointly optimizes the
frequency-domain precoding, subcarrier user selection, and
power allocation on the downlink of  coordinated multicell OFDMA systems. In addition, authors in \cite{Suh2008} consider the uplink of a limited-size cellular system without BS cooperation, showing that the interference-free dofs can be achieved as the number of UTs grows. Employing IA with unit multiplexing gain towards the non-intended BSs, they study the effect of multi-path channels and single-path channels with propagation delay. Furthermore, the concept of decomposable channel is employed to enable a modified scheme called subspace IA, which is able to simultaneously align interference towards multiple non-intended receivers over a multidimensional space.
Finally, the effect of limited feedback on cellular IA schemes has been investigated and quantified in \cite{Tresch2009b,Lee2010}.
\section{Channel Model \& Throughput Analysis}
\label{sec: eigenvalue distribution analysis}

In this paper, the considered system comprises a modified version of Wyner's linear cellular array \cite{Wyner1994,Somekh2007a,Somekh2007_Chapter}, which has been used extensively as a tractable model for studying MJD scenarios. In the modified model studied herein, MJD is possible for clusters of $M$ adjacent BSs while the focus is on the uplink. Unlike  \cite{Suh2008,Tresch2009}, IA is employed herein to mitigate intercluster interference between cluster-edge cells. Let us assume that $K$ UTs are positioned between each pair of neighboring BSs with path loss coefficients 1 and $\alpha$ respectively (Figure \ref{fig: cellular array}). All BSs and UTs are equipped with $n=K+1$ antennas to enable IA over the multiple spatial dimensions for the clustered UTs. 
In this setting, four scenarios of intercluster interference are considered, namely global MJD, IA, RDMA and CI. It should be noted that only cluster-edge UTs employ interference mitigation techniques, while UTs in the interior of the cluster use the optimal wideband transmission scheme with superposition coding as in \cite{Somekh2000}. Successive interference cancellation is employed in each cluster processor in order to recover the UT signals. Furthermore, each cluster processor has full CSI for all the wireless links in its coverage area. The following subsections explain the mode of operation for each approach and describe the analytical derivation of the per-cell sum-rate throughput. 

\subsection{Global Multicell Joint Decoding}
In global MJD, a central processor is able to jointly decode the signals received by neighboring clusters and therefore no intercluster interference takes place. In other words, the entire cellular system can be assumed to be comprised of a single extensive cluster. As it can be seen, this case serves as an upper bound to the IA case. The received $n\times 1$ symbol vector $\mathbf{y}_i$ at any random BS can be expressed as follows:
\begin{equation}
\mathbf{y}_i(t)=\mathbf{G}_{i,i}(t)\mathbf{x}_i(t)+\alpha\mathbf{G}_{i,i+1}(t)\mathbf{x}_{i+1}(t)+\mathbf{z}_i(t),
\label{eq: full cooperation channel model}
\end{equation}
where the $n\times 1$ vector $\mathbf{z}$ denotes AWGN with $\mathbb{E}[\mathbf{z}_i]=\mathbf{0}$ and $\mathbb{E}[\mathbf{z}_i\mathbf{z}_i^H]=\mathbf{I}$. The $Kn\times 1$ vector $\mathbf{x}_i$ denotes the transmitted symbol vector of the $i$th UT group with $\mathbb{E}[\mathbf{x}_i\mathbf{x}_i^H]=\gamma\mathbf{I}$ where $\gamma$ is the transmit Signal to Noise Ratio per UT antenna. The $n\times Kn$ channel matrix $\mathbf{G}_{i,i}\sim\mathcal{CN}(\mathbf{0},\mathbf{I}_n)$ includes the flat fading coefficients of the $i$th UT group towards the $i$th BS\ modelled as independent identically distributed (i.i.d.) complex circularly symmetric (c.c.s.) random variables. Similarly, the term $\alpha\mathbf{G}_{i,i+1}(t)\mathbf{x}_{i+1}(t)$ represents the received signal at the $i$th BS originating from the UTs of the neighboring cell indexed $i+1$. The scaling factor \(\alpha<1\) models the amount of received intercell interference which depends on the path loss model and the density of the cellular system\footnote{For more details on the modelling of the \(\alpha\) parameter, the reader is referred to \cite{Chatzinotas_Chapter1}.}. Another intuitive description of the \(\alpha\) factor is that it models the power imbalance between intra-cell and inter-cell signals. 

Assuming a memoryless channel, the system channel model can be written in a vectorial form as follows:
\begin{equation}
\mathbf{y}=\mathbf{H}\mathbf{x}+\mathbf{z},
\end{equation}
where the aggregate channel matrix has dimensions $Mn\times (M+1)Kn$ and can be modelled as:
\begin{equation}
\mathbf{H}=\mathbf{\Sigma} \odot \mathbf{G}
\label{eq: GMJD channel matrix}
\end{equation}
with $\mathbf{\Sigma}=\mathbf{\tilde \Sigma}\otimes\mathbb{I}_{n\times Kn}$ being a block-Toeplitz matrix and $\mathbf{G}\sim\mathcal{CN}(\mathbf{0},\mathbf{I}_{Mn})$. In addition,  $\mathbf{\tilde \Sigma}$ is a $M\times M+1$ Toeplitz matrix structured as follows:
\begin{equation}
\mathbf{\tilde \Sigma}=
 \left[
  \begin{array}{ c c c c c}
     1 & \alpha & 0 & \cdots & 0 \\
     0 & 1 & \alpha &  & 0 \\
     \vdots &  & \ddots & \ddots & \vdots \\
     0 & \cdots & 0 & 1 & \alpha 
  \end{array} 
  \right]
\end{equation}
Assuming no CSI at the UTs, the per-cell capacity is given by the MIMO Multiple Access (MAC) channel capacity:
\begin{align}
\mathrm{C}_\mathrm{MJD}&=\frac{1}{M}\mathbb{E}\left[\mathcal{I}\left( \mathbf{x};\mathbf{y}\mid \mathbf{H}\right)\right]\nonumber
\\&=\frac{1}{M}\mathbb{E}\left[\log\det\left(\mathbf{I}_{Mn}+\gamma\mathbf{HH}^H\right)\right].
\label{eq: global MJD simul}
\end{align}

\begin{thm}
In the global MJD case, the per-cell capacity for asymptotically large \(n\) converges almost surely (a.s.) to the Mar\v cenko-Pastur (MP) law with appropriate scaling \cite{Chatzinotas_letter,Chatzinotas_JWCOM}:
\label{thm: global MJD}
\begin{equation}
\mathrm{C}_\mathrm{MJD}\stackrel{_{a.s.}}{\longrightarrow}Kn\mathcal{V}_{\mathrm{MP}}\left(\frac{M}{M+1}n\gamma\left(1+\alpha^2\right),K\frac{M+1}{M}\right),
\label{eq: full cooperation capacity}
\end{equation}
\begin{align}
\text{where\ \ \ }\mathcal{V}_{\mathrm{MP}}\left(\gamma,\beta\right)&=\mathrm{log}\left( 1+\gamma-\frac{1}{4}\phi\left( \gamma,\beta \right)\ \right)+\frac{1}{\beta}\mathrm{log}\left( 1+\gamma \beta-\frac{1}{4}\phi\left( \gamma,\beta \right)\ \right) -\frac{1}{4\beta
\gamma}\phi\left( \gamma,\beta\right)\nonumber
\\\text{and\ \ \ }\phi\left( \gamma,\beta\right)&=\left(\sqrt{\gamma\left( 1+\sqrt{\beta} \right)^2+1}-\sqrt{\gamma\left( 1-\sqrt{\beta}\right)^2+1}\right)^{2}.\nonumber
\end{align}
\begin{proof}
For the sake of completeness and to facilitate latter derivations, an outline of the proof in \cite{Chatzinotas_letter,Chatzinotas_JWCOM} is provided here. The derivation of this expression is based on an asymptotic analysis in the number of antennas $n\rightarrow\infty$:
\begin{align}
\frac{1}{n}C_\mathrm{MJD}
&=\lim_{n\rightarrow\infty}\frac{1}{Mn}\mathbb{E}\left[\log \det \left( \mathbf{I}_{Mn}+\gamma\mathbf{H}\mathbf{H}^{H}\right) \right]\nonumber
\\
&=\lim_{n\rightarrow\infty}\mathbb{E}\left[\frac{1}{Mn}\sum_{i=1} ^{Mn}\log\left( 1+M\tilde\gamma\lambda _{i}\left(\frac{1}{Mn}\mathbf{H}\mathbf{H}^{H}\right)\right) \right]\nonumber
\\&=\int_0^\infty\log\left( 1+M\tilde\gamma x \right)f^{\infty}_{\frac{1}{Mn}\mathbf{H}\mathbf{H}^H}(x)dx\nonumber
\\&=K\int_0^\infty\log\left( 1+M\tilde\gamma x \right)f^{\infty}_{\frac{1}{Mn}\mathbf{H}^H\mathbf{H}}(x)dx\nonumber
\\&=K\mathcal{V}_{\frac{1}{Mn}\mathbf{H}^H\mathbf{H}}(M\tilde\gamma)\nonumber
\\&\stackrel{_{a.s.}}{\longrightarrow}K\mathcal{V}_{\mathrm{MP}}(q\left(\mathbf{\Sigma}\right)M\tilde\gamma,K),
\label{eq: full cooperation capacity proof}
\end{align}
where \(\lambda_{i}\left(\mathbf{X}\right)\) and $f^\infty_\mathbf{X}$ denote the eigenvalues and the asymptotic eigenvalue probability distribution function (a.e.p.d.f.) of matrix \(\mathbf{X}\) respectively and \(\mathcal{V}_{\mathbf{X}}(x)=\mathbb{E}[\log(1+x\mathbf{X})]\) denotes the Shannon transform of \(\mathbf{X}\) with scalar parameter \(x\). It should be noted that \(\tilde \gamma=n\gamma\) denotes the total UT transmit power normalized by the receiver noise power\footnote{For the purposes of the analysis the variable $\tilde \gamma$ is kept finite as the number of antennas $Mn$ grows large, so that the system power does not grow to infinity.}. The last step of the derivation is based on unit rank matrices decomposition and analysis on the R-transform domain, as presented in \cite{Chatzinotas_letter,Chatzinotas_JWCOM}. The scaling factor
\begin{equation}
q(\mathbf{\Sigma})\triangleq\left\Vert \mathbf{\Sigma}\right\Vert^2/\left(Mn\times (M+1)Kn\right)
\label{eq: q summation}
\end{equation}
is the Frobenius norm of the \(\mathbf{\Sigma}\) matrix \(\left\Vert \mathbf{\Sigma}\right\Vert\triangleq\sqrt{tr\left\{ \mathbf{\Sigma}^H\mathbf{\Sigma}\right\}}\) normalized by the matrix dimensions and 
\begin{align}
q\left(\mathbf{\Sigma}\right)\stackrel{_{(a)}}{=}q\left(\mathbf{\tilde \Sigma}\right)=\frac{1+\alpha^2}{M+1}
\label{eq:variance profile matrix norm}
\end{align} 
where step $(a)$ follows from \cite[Eq.(34)]{Chatzinotas_JWCOM}.
\end{proof}
\end{thm}

\subsection{Interference Alignment}
In order to evaluate the effect of IA as an intercluster interference mitigation technique, a simple precoding scheme is assumed for the cluster-edge UT groups, inspired by \cite{Suh2008}. Let us assume a $n\times 1$ unit norm reference vector $\mathbf{v}$ with $\Vert \mathbf v\Vert^2=n$ and 
\begin{equation}
\mathbf{y}_1=\mathbf{G}_{1,1}\mathbf{x}_1+\alpha\mathbf{G}_{1,2}\mathbf{x}_{2}+\mathbf{z}_{1}
\label{eq: first BS signal}
\end{equation}
\begin{equation}
\mathbf{y}_M=\mathbf{G}_{M,M}\mathbf{x}_M+\alpha\mathbf{G}_{M,M+1}\mathbf{x}_{M+1}+\mathbf{z}_{M},
\end{equation}
where $\mathbf{y}_1$ and $\mathbf{y}_M$ represent the received signal vectors at the first and last BS of the cluster respectively. The first UT group has to align its input \(\mathbf x_1\) towards the non-intended BS of the cluster on the left (see Fig. \ref{fig: cellular array}), while the \(M\)th BS has to filter our the aligned interference coming from the \(M+1\)th UT group which belongs to the cluster on the right. These two strategies are described in detail in the following subsections:
\subsubsection{Aligned Interference Filtering}
The objective is to suppress the term $\alpha\mathbf{G}_{M,M+1}\mathbf{x}_{M+1}$ which represents intercluster interference. It should be noted that UTs of the $M+1$th cell are assumed to have perfect CSI about the channel coefficients $\mathbf{G}_{M,M+1}$. Let us also assume that $\mathbf{x}^j_i$ and $\mathbf{G}^j_{\tilde i,i}$ represent the transmitter vector and channel matrix  of the $j$th UT in the $i$th group towards the \(\tilde i\)th BS. In this context, the following precoding scheme is employed to align interference:
\begin{equation}
\mathbf{x}^j_{M+1}=\left(\mathbf{G}^j_{M,M+1}\right)^{-1}\mathbf{v}_jx^j_{M+1},
\end{equation}
where \(\mathbf v_j=\mathbf vv_j\) is a scaled version of \(\mathbf v\) which satisfies the input power constraint \(\mathbb{E}[\mathbf{x}^j_{M+1}{\mathbf{x}^j_{M+1}}^H]=\gamma\mathbf{I}\).
This precoding results in unit multiplexing gain and is by no means the optimal IA scheme, but it serves as a tractable way of evaluating the IA performance. %The extra receiver dimensions ensures the feasibility of IA.
Following this approach, the intercluster interference can be expressed as:
\begin{equation}
\alpha\mathbf{G}_{M,M+1}\mathbf{x}_{M+1}=\alpha\sum_{j=1}^K\mathbf{G}^j_{M,M+1}\mathbf{x}^j_{M+1}=\alpha\sum_{j=1}^K\mathbf{G}^j_{M,M+1}\left(\mathbf{G}^j_{M,M+1}\right)^{-1}\mathbf{v}v_jx^j_{M+1}=\alpha\mathbf{v}\sum_{j=1}^Kv_jx^j_{M+1}.
\end{equation}
It can be easily seen that interference has been aligned across the reference vector and it can be removed using a $K\times n$ zero-forcing filter $\mathbf{Q}$ designed so that $\mathbf Q$ is a truncated unitary matrix \cite{Cadambe2009} and $\mathbf{Qv}=\mathbf{0}$. After filtering, the received signal at the $M$th BS can be expressed as:
\begin{equation}
\mathbf{\tilde y}_M=\mathbf{QG}_{M,M}\mathbf{x}_M+\mathbf{\tilde z}_{M},
\label{eq: IA filtering}
\end{equation}
Assuming that the system operates in high-SNR regime and  is therefore interference limited, the effect of the AWGN noise colouring $\mathbf{\tilde z}_{M}=\mathbf{Qz}_{M}$ can be ignored, namely $\mathbb{E}[\mathbf{\tilde z}_{M}\mathbf{\tilde z}_{M}^H]=\mathbf I_{K}$. 
\begin{lem}
The Shannon transform of the covariance matrix of $\mathbf{QG}_{M,M}$ is equivalent to that of a $K\times K$ Gaussian matrix $\mathbf{G}_{K\times K}$. 
\label{lem: IA gaussian}
\begin{proof}
Using the property $\det(\mathbf{I+\gamma AB})=\det(\mathbf{I+\gamma BA})$, it can be written that:
\begin{align}
\det\left(\mathbf{I}_{K}+\gamma \mathbf{QG}_{M,M}(\mathbf{QG}_{M,M})^H\right)&=\det\left(\mathbf{I}_{K}+\gamma \mathbf{QG}_{M,M}\mathbf{G}_{M,M}^H\mathbf{Q}^H\right)\nonumber
\\&=\det\left(\mathbf{I}_{n}+\gamma \mathbf{G}_{M,M}^H\mathbf{Q}^H\mathbf{Q}\mathbf{G}_{M,M}\right).
\label{eq: det derivation}
\end{align}
The $K\times n$ truncated unitary matrix $\mathbf{Q}$ has $K$ unit singular values and therefore the matrix product  $\mathbf{Q}^H\mathbf{Q}$ has $K$ unit eigenvalues and a zero eigenvalue. Applying eigenvalue decomposition on $\mathbf{Q}^H\mathbf{Q}$, the left and right eigenvectors can be absorbed
by the isotropic Gaussian matrices $\mathbf{G}_{M,M}^H$ and $\mathbf{G}_{M,M}$ respectively, while the zero eigenvalue removes one of the $n$ dimensions.
Using the definition of Shannon transform \cite{Tulino04}, eq. \eqref{eq: det derivation} yields
\begin{equation}
\mathcal{V}_{\mathbf{QG}_{M,M}(\mathbf{QG}_{M,M})^H}(\gamma)=\mathcal{V}_{\mathbf{G}_{K\times K}\mathbf{G}_{K\times K}^H}(\gamma).
\end{equation}
\end{proof} 
\end{lem}
Based on this lemma and for the purposes of the analysis, \(\mathbf{QG}_{M,M}\) is replaced by $\mathbf{G}_{K\times K}$ in the equivalent channel matrix.
\subsubsection{Interference Alignment}
\begin{figure}
        \centering
                \includegraphics[width=0.8\textwidth]{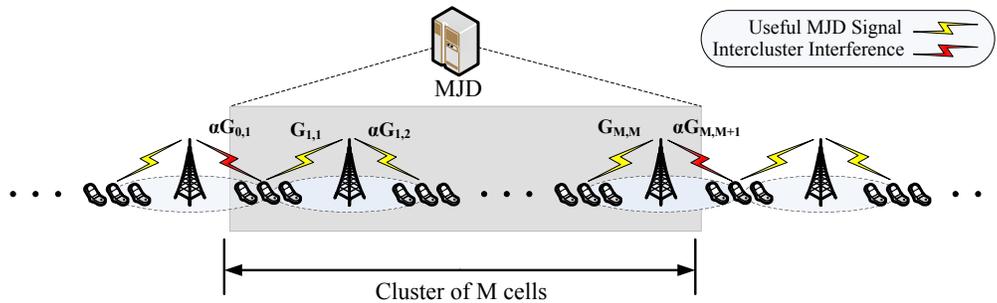}
        \caption{Graphical representation of the considered cellular system modelled as a modified version of Wyner's model. $K$ UTs are positioned between each pair of neighboring BSs with path loss coefficients 1 and $\alpha$ respectively. All BSs and UTs are equipped with $n=K+1$ antennas. The UTs positioned within the box shall be jointly processed. The red links denote intercluster interference.}
        \label{fig: cellular array}
\end{figure}

The $M$th BS has filtered out incoming interference from the cluster on the right (Figure \ref{fig: cellular array}), but outgoing  intercluster interference should be also aligned to complete the analysis. This affects the first UT group which should align its interference towards the $M$th BS of the cluster on the left (Figure \ref{fig: cellular array}). Following the same precoding scheme and using eq. \eqref{eq: first BS signal}
\begin{equation}
\mathbf{G}_{1,1}\mathbf{x}_{1}=\sum_{j=1}^K\mathbf{G}^j_{1,1}\mathbf{x}^j_{1}=\sum_{j=1}^K\mathbf{G}^j_{1,1}\left(\mathbf{G}^j_{0,1}\right)^{-1}\mathbf{v}v_jx^j_{1},
\end{equation}
where $\mathbf{G}^j_{0,1}$ represents the fading coefficients of the $j$th UT of the first group towards the $M$th BS of the neighboring cluster on the left. Since the exact eigenvalue distribution of the matrix product $\mathbf{G}^j_{1,1}(\mathbf{G}^j_{0,1})^{-1}\mathbf{v}v_j$ is not straightforward to derive, for the purposes of rate analysis it is approximated by a Gaussian vector with unit variance. This approximation implies that IA precoding does not affect the statistics of the equivalent channel towards the intended BS. 
\subsubsection{Equivalent Channel Matrix}
To summarize, IA has the following effects on the channel matrix $\mathbf H$ used for the case of global MJD. The intercluster interference originating from the $M+1$th UT group is filtered out and thus $Kn$ vertical dimensions are lost. During this process, one horizontal dimension of the $M$th BS is also filtered out, since it contains the aligned interference from the $M+1$th UT group. Finally, the first UT group has to precode in order to align its interference towards the $M$th BS of neighboring cluster and as a result only $K$ out of $Kn$ dimensions are preserved.
% with an array gain of $\sqrt{n}$. 
The resulting channel matrix can be described as follows:
\begin{equation}
\mathbf{H}_\mathrm{IA}=\mathbf{\Sigma}_\mathrm{IA} \odot \mathbf{G}_\mathrm{IA},
\label{eq: IA channel matrix}
\end{equation}
where $\mathbf{G}_\mathrm{IA}\sim\mathcal{CN}(\mathbf{0},\mathbf{I}_{Mn-1})$ and
\begin{align}
\mathbf{\Sigma}_\mathrm{IA}=\begin{bmatrix}
\mathbf{\Sigma}_1 \\
\mathbf{\Sigma}_2 \\
\mathbf{\Sigma}_3 \\
\end{bmatrix}
\label{eq: three sigmas}
\end{align}
with $\mathbf{\Sigma}_1=[\mathbb{I}_{n\times K}\ \alpha\mathbb{I}_{n\times Kn}\ \mathbf{0}_{n\times (M-2)Kn}]$ being a $n\times (M-1)Kn+K$ matrix\footnote{The structure of the first block of $\mathbf{\Sigma}_1$ originates in the Gaussian approximation of $\frac{1}{\alpha}\mathbf{G}^j_{1,1}(\mathbf{G}^j_{0,1})^{-1}\mathbf{v}v_j$.},
$\mathbf{\Sigma}_2=[\mathbf{0}_{(M-2)n\times K}\ \mathbf{\tilde \Sigma}_{M-2\times M-1}\otimes \mathbb{I}_{n\times Kn}]$ being a $(M-2)n\times (M-1)Kn+K$ matrix
and $\mathbf{\Sigma}_3=[\mathbf{0}_{n-1\times (M-2)Kn+K}\ \mathbb{I}_{n-1\times Kn}]$ being a $n-1\times (M-1)Kn+K$ matrix\footnote{The structure of the last block of $\mathbf{\Sigma}_3$ is based on Lemma \ref{lem: IA gaussian}.}.

Since all intercluster interference has been filtered out and the effect of filter \(\mathbf Q\) has been already incorporated in the structure of \(\mathbf{H}_\mathrm{IA}\), the per-cell throughput in the IA case is still given by the MIMO MAC expression:
\begin{align}
\mathrm{C}_\mathrm{IA}&=\frac{1}{M}\mathbb{E}\left[\mathcal{I}\left( \mathbf{x};\mathbf{y}\mid \mathbf{H}_\mathrm{IA}\ \right)\right]
\label{eq: IA simul}
\\&=\frac{1}{M}\mathbb{E}\left[\log\det\left(\mathbf{I}_{Mn-1}+\gamma\mathbf{H}_\mathrm{IA}\mathbf{H}_\mathrm{IA}^H\right)\right].\nonumber
\end{align}

\begin{thm}
In the IA case, the per-cell throughput can be derived from the R-transform of the a.e.p.d.f. of matrix ${\frac{1}{n}\mathbf{H}_\mathrm{IA}^H\mathbf{H}_\mathrm{IA}}$.
\label{thm: IA th 1}
\begin{proof}
Following an asymptotic analysis where $n\rightarrow\infty$:
\begin{align}
\frac{1}{n}C_\mathrm{IA}
&=\lim_{n\rightarrow\infty}\frac{1}{Mn}\mathbb{E}\left[\log \det \left( \mathbf{I}_{Mn-1}+\gamma \mathbf{H}_\mathrm{IA}\mathbf{H}_\mathrm{IA}^{H}\right) \right]\nonumber
\\
&=\frac{Mn-1}{Mn}\lim_{n\rightarrow\infty}\mathbb{E}\left[\frac{1}{Mn-1}\sum_{i=1} ^{Mn-1}\log\left( 1+\tilde\gamma\lambda _{i}\left(\frac{1}{n}\mathbf{H}_\mathrm{IA}\mathbf{H}_\mathrm{IA}^{H}\right)\right) \right]\nonumber
\\&=\frac{Mn-1}{Mn}\int_0^\infty\log\left( 1+\tilde\gamma x \right)f^{\infty}_{\frac{1}{n}\mathbf{H}_\mathrm{IA}\mathbf{H}_{\mathrm{IA}}^H}(x)dx\nonumber
\\&=\frac{(M-1)Kn+K}{Mn}\int_0^\infty\log\left( 1+\tilde\gamma x \right)f^{\infty}_{\frac{1}{n}\mathbf{H}_\mathrm{IA}^H\mathbf{H}_\mathrm{IA}}(x)dx\label{eq: interference alignment capacity}
\end{align}
The a.e.p.d.f. of $\frac{1}{n}\mathbf{H}_\mathrm{IA}^H\mathbf{H}_\mathrm{IA}$ is obtained by determining
the imaginary part of the Stieltjes transform \(\mathcal{S}\) for real arguments
\begin{equation}
f^{\infty}_{\frac{1}{n}\mathbf{H}_\mathrm{IA}^H\mathbf{H}_\mathrm{IA}}(x)=\lim_{y\rightarrow0^+}\frac{1}{\pi}\mathfrak{I}\left\{ \mathcal{S}_{\frac{1}{n}\mathbf{H}_\mathrm{IA}^H\mathbf{H}_\mathrm{IA}}(x+jy)\ \right\}
\label{eq: limiting eigenvalue pdf}
\end{equation}
considering that the Stieltjes transform is derived from the R-transform
\cite{Rao2007} as follows 
\begin{equation}
\mathcal{S}_{\frac{1}{n}\mathbf{H}_\mathrm{IA}^H\mathbf{H}_\mathrm{IA}}^{-1}(z)=\mathcal{R}_{\frac{1}{n}\mathbf{H}_\mathrm{IA}^H\mathbf{H}_\mathrm{IA}}(-z)-\frac{1}{z}.
\label{eq: Cauchy}
\end{equation}
\end{proof}
\end{thm}

\begin{thm}
The R-transform of the a.e.p.d.f. of matrix ${\frac{1}{n}\mathbf{H}_\mathrm{IA}^H\mathbf{H}_\mathrm{IA}}$ is given by:
\begin{equation}
\mathcal{R}_{\frac{1}{n}\mathbf{H}_\mathrm{IA}^H\mathbf{H}_\mathrm{IA}}(z)=\sum_{i=1}^3\mathcal{R}_{\frac{1}{n}\mathbf{H}_i^H\mathbf{H}_i}(z,k_i,\beta_i,q_i)
\label{eq: interference alignment R-transform}
\end{equation}
with $k,\beta,q$ parameters given by:
\begin{align}
\mathbf{H}_1:&k_1=\frac{K+2}{MK+M-K},\beta_1=\frac{K}{K+1}+K,q_1=\frac{1+(K+1)\alpha^2}{K+2}\nonumber
\\\mathbf{H}_2:&k_2=\frac{(M-1)(K+1)}{MK+M-K},\beta_2=\frac{M-1}{M-2}K,q_2=\frac{M-2}{M-1}(1+\alpha^2)\nonumber
\\\mathbf{H}_3:&k_3=\frac{K+1}{MK+M-K},\beta_3=K+1,q_3=1\nonumber
\end{align}
and \(\mathcal{R}_{\frac{1}{n}\mathbf{H}_i^H\mathbf{H}_i}\) given by theorem \ref{thm: concatenated R transform}. 
\label{thm: IA th 2}
\begin{proof}
Based on eq.\eqref{eq: three sigmas}, the matrix $\mathbf{H}_\mathrm{IA}^H\mathbf{H}_\mathrm{IA}$ can be decomposed as the following sum:
\begin{equation}
\mathbf{H}_\mathrm{IA}^H\mathbf{H}_\mathrm{IA}=\mathbf{H}_1^H\mathbf{H}_1+\mathbf{H}_2^H\mathbf{H}_2+\mathbf{H}_3^H\mathbf{H}_3, 
\end{equation}
where $\mathbf{H}_1=\mathbf{\Sigma}_1\odot \mathbf{G}_{n\times (M-1)Kn+K}$, $\mathbf{H}_2=\mathbf{\Sigma}_2\odot \mathbf{G}_{(M-2)n\times (M-1)Kn+K}$ and $\mathbf{H}_3=\mathbf{\Sigma}_3\odot \mathbf{G}_{n-1\times (M-1)Kn+K}$.
Using the property of free additive convolution \cite{Tulino04} and Theorem \ref{thm: concatenated R transform} in Appendix \ref{app: 1}, eq. \eqref{eq: interference alignment R-transform} holds in the R-transform domain.
\end{proof}
\end{thm}

\subsection{Resource Division Multiple Access}
RDMA entails that the available time or frequency resources are divided into two orthogonal parts assigned to cluster-edge cells in order to eliminate intercluster interference \cite[Efficient isolation scheme]{Katranaras2009}. While double the power can be transmitted in each orthogonal part, the available channel resources are cut in half for cluster-edge cells. The channel modelling is similar to the one in global MJD case (eq. \eqref{eq: full cooperation channel model}), although in this case the throughput is analyzed separately for each orthogonal part and subsequently averaged. 
Assuming no CSI at the UTs, the per-cell throughput in the RDMA case is given by:
\begin{align}
\mathrm{C}_\mathrm{RD}&=\frac{\mathrm{C}_{\mathrm{RD}_1}+\mathrm{C}_{\mathrm{RD}_2}}{2}\nonumber
\\&=\frac{1}{2M}\left(\mathbb{E}\left[\log \det \left( \mathbf{I}_{Mn}+\gamma \mathbf{H}_\mathrm{RD_1}\mathbf{H}_\mathrm{RD_1}^{H}\right) \right]+\mathbb{E}\left[\log \det \left( \mathbf{I}_{(M-1)n}+\gamma \mathbf{H}_\mathrm{RD_2}\mathbf{H}_\mathrm{RD_2}^{H}\right) \right]\right),
\label{eq: RDMA simul}
\end{align}
where $\mathrm{C}_{\mathrm{RD}_1}$ and $\mathrm{C}_{\mathrm{RD}_2}$ denote the capacities when the cluster-edge UTs are active and inactive respectively.
When the cluster-edge UTs are active, the cluster processor receives signals from all $M$ BSs and the resulting $Mn\times MKn$ channel matrix is structured as follows:
\begin{align}
\mathbf{H}_{\mathrm{RD}_1}&=\mathbf{\Sigma}_{\mathrm{RD}_1}\odot\mathbf{G}_{Mn\times MKn}\text{\ \ with\ \ }\mathbf{H}_{\mathrm{RD}_1}=\begin{bmatrix}
\mathbf{\tilde H} \\
\mathbf{\tilde H}_{\mathrm{RD}_1} \\
\end{bmatrix}\nonumber
\\\mathbf{\Sigma}_{\mathrm{RD}_1}&=\begin{bmatrix}
\mathbf{\tilde \Sigma} \\
\mathbf{\tilde \Sigma}_{\mathrm{RD}_1} \\
\end{bmatrix}
\text{\ \ and\ \ } \mathbf{\tilde \Sigma}_{\mathrm{RD}_1}=[\mathbf{0}_{n\times (M-1)Kn}\ 2\mathbb{I}_{n\times Kn}],
\label{eq: RDMA matrix decomposition}
\end{align}
where the factor $2$ is due to the doubling of the transmitted power. 
\begin{thm}
In the RDMA case with active cluster-edge UTs, the per-cell throughput $\mathrm{C}_{\mathrm{RD}_1}$ can be derived from the R-transform of the a.e.p.d.f. of matrix ${\frac{1}{n}\mathbf{H}_{\mathrm{RD}_1}^H\mathbf{H}_{\mathrm{RD}_1}}$, where:
\begin{equation}
\mathcal{R}_{\frac{1}{n}\mathbf{H}_{\mathrm{RD}_1}^H\mathbf{H}_{\mathrm{RD}_1}}(z)=\frac{({M-1})(1+\alpha^2)}{1-(1+\alpha^2)K{M}z}+\mathcal{R}_{\frac{1}{n}\mathbf{B}^H\mathbf{B}}(z,\frac{1}{M},K,2).
\label{eq: active R-transform}
\end{equation}
\label{thm: active RDMA}
\begin{proof}
Following an asymptotic analysis where $n\rightarrow\infty$:
\begin{align}
\frac{1}{n}C_\mathrm{RD_1}
&=\lim_{n\rightarrow\infty}\frac{1}{Mn}\mathbb{E}\left[\log \det \left( \mathbf{I}_{Mn}+\gamma \mathbf{H}_\mathrm{RD_1}\mathbf{H}_\mathrm{RD_1}^{H}\right) \right]\nonumber
\\&=K\int_0^\infty\log\left( 1+\tilde\gamma x \right)f^{\infty}_{\frac{1}{n}\mathbf{H}_\mathrm{RD_1}^H\mathbf{H}_\mathrm{RD_1}}(x)dx.
\end{align}
Using the matrix decomposition of eq. \eqref{eq: RDMA matrix decomposition} and free additive convolution \cite{Tulino04}:
\begin{equation}
\mathcal{R}_{\frac{1}{n}\mathbf{H}_{\mathrm{RD}_1}^H\mathbf{H}_{\mathrm{RD}_1}}(z)=\mathcal{R}_{\frac{1}{n}\mathbf{\tilde H}^H\mathbf{\tilde H}}(z)+\mathcal{R}_{\frac{1}{n}\mathbf{\tilde H}_{\mathrm{RD}_1}^H\mathbf{\tilde H}_{\mathrm{RD}_1}}(z).
\end{equation}
Eq. \eqref{eq: active R-transform} follows from eq. \eqref{eq: variance R-transform} with $q=Mq(\mathbf{\tilde \Sigma})=({M-1})(1+\alpha^2),\beta=KM/(M-1)$ and theorem \ref{thm: concatenated R transform}.
\end{proof}
\end{thm}

When the cluster-edge UTs are inactive, the cluster processor receives signals from  $M-1$ BSs and the resulting $M-1\times (M-1)Kn$ channel matrix is structured as follows:
\begin{align}
\mathbf{H}_{\mathrm{RD}_2}&=\mathbf{\Sigma}_{\mathrm{RD}_2}\odot\mathbf{G}_{(M-1)n\times (M-1)Kn}\text{\ \ with\ \ }\mathbf{H}_{\mathrm{RD}_2}=\begin{bmatrix}
\mathbf{\tilde H} \\
\mathbf{\tilde H}_{\mathrm{RD}_2} \\
\end{bmatrix}\nonumber
\\\mathbf{\Sigma}_{\mathrm{RD}_2}&=\begin{bmatrix}
\mathbf{\tilde \Sigma} \\
\mathbf{\tilde \Sigma}_{\mathrm{RD}_2} \\
\end{bmatrix}
\text{\ \ and\ \ } \mathbf{\tilde \Sigma}_{\mathrm{RD}_2}=[\mathbf{0}_{n\times (M-2)Kn}\ \mathbb{I}_{n\times Kn}].
\label{eq: RDMA matrix decomposition 2}
\end{align}
\begin{thm}
In the RDMA case with inactive cluster-edge UTs, the per-cell throughput $\mathrm{C}_{\mathrm{RD}_2}$ can be derived from the R-transform of the a.e.p.d.f. of matrix ${\frac{1}{n}\mathbf{H}_{\mathrm{RD}_2}^H\mathbf{H}_{\mathrm{RD}_2}}$, where:
\begin{equation}
\mathcal{R}_{\frac{1}{n}\mathbf{H}_{\mathrm{RD}_2}^H\mathbf{H}_{\mathrm{RD}_2}}(z)=\frac{({M-2})(1+\alpha^2)}{1-(1+\alpha^2)K({M-1})z}+\mathcal{R}_{\frac{1}{n}\mathbf{B}^H\mathbf{B}}(z,\frac{1}{M-1},K,1)
\label{eq: active R-transform 2}
\end{equation}
\label{thm: inactive RDMA}
\begin{proof}
Following an asymptotic analysis where $n\rightarrow\infty$:
\begin{align}
\frac{1}{n}C_\mathrm{RD_2}
&=\lim_{n\rightarrow\infty}\frac{1}{Mn}\mathbb{E}\left[\log \det \left( \mathbf{I}_{(M-1)n}+\gamma \mathbf{H}_\mathrm{RD_2}\mathbf{H}_\mathrm{RD_2}^{H}\right) \right]\nonumber
\\&=K\frac{M-1}{M}\int_0^\infty\log\left( 1+\tilde\gamma x \right)f^{\infty}_{\frac{1}{n}\mathbf{H}_\mathrm{RD_2}^H\mathbf{H}_\mathrm{RD_2}}(x)dx.
\end{align}The rest of this proof follows the steps of Theorem \ref{thm: active RDMA}.
\end{proof}
\end{thm}

\subsection{Cochannel Interference Allowance}
CI is considered as a worst case scenario where no signal processing is performed in order to mitigate intercluster interference \cite{ChatzinotasArxiv2010}. As it can be seen, this case serves as a lower bound to the IA case. The channel modelling is identical with the one in global MJD case (eq. \eqref{eq: full cooperation channel model}), although in this case the cluster-edge UT group contribution $\alpha\mathbf{G}_{M,M+1}(t)\mathbf{x}_{M+1}(t)$ is considered as interference. As a result, the interference channel matrix can be expressed as:
\begin{equation}
\mathbf{H}_\mathrm{I}=
\begin{bmatrix}
\mathbf{0}_{Mn\times Kn} \\
\alpha\mathbf{G}_{n\times Kn} \\
\end{bmatrix}.
\label{eq: CI channel matrix}
\end{equation}
Assuming no CSI at the UTs, the per-cell throughput in the CI case is given by:
\begin{equation}
\mathrm{C}_\mathrm{CI}=\mathrm{C}_\mathrm{MJD}-\mathrm{C}_\mathrm{I},
\label{eq: cia simul 1}
\end{equation}
where $\mathrm{C}_\mathrm{I}$ denotes the throughput of the interfering UT group normalized by the cluster size:
\begin{align}
\mathrm{C}_\mathrm{I}&=\frac{1}{M}\mathbb{E}\left[\mathcal{I}\left( \mathbf{x};\mathbf{y}\mid \mathbf{H}_\mathrm{I}\ \right)\right]\nonumber
\\&=\frac{1}{M}\mathbb{E}\left[\log\det\left(\mathbf{I}_{Mn}+\gamma\mathbf{H}_\mathrm{I}\mathbf{H}_\mathrm{I}^H\right)\right].
\label{eq: cia simul 2}
\end{align}
\begin{thm}
In the CI case, the per-cell throughput converges almost surely (a.s.) to a difference of two scaled versions of the the MP law:
\begin{equation}
C_\mathrm{CI}\stackrel{_{a.s.}}{\longrightarrow}Kn\mathcal{V}_{\mathrm{MP}}\left(\frac{M}{M+1}n\gamma\left(1+\alpha^2\right),K\frac{M+1}{M}\right)-\frac{Kn}{M}\mathcal{V}_{\mathrm{MP}}(\alpha^2 n\gamma,K).
\label{eq: cochannel interference capacity}
\end{equation}
\label{thm: cia thm}
\begin{proof}
Following an asymptotic analysis in the number of antennas $n\rightarrow\infty$::
\begin{align}
\frac{1}{n}C_\mathrm{I}
&=\lim_{n\rightarrow\infty}\frac{1}{Mn}\mathbb{E}\left[\log \det \left( \mathbf{I}_{Mn}+\gamma \mathbf{H}_\mathrm{I}\mathbf{H}_\mathrm{I}^{H}\right) \right]\nonumber
\\
&=\lim_{n\rightarrow\infty}\frac{1}{Mn}\mathbb{E}\left[\log \det \left( \mathbf{I}_{n}+\gamma\alpha^2\mathbf{G}_{n\times Kn}\mathbf{G}_{n\times Kn}^{H}\right) \right]\nonumber
\\&\stackrel{_{a.s.}}{\longrightarrow}\frac{K}{M}\mathcal{V}_{\mathrm{MP}}(\alpha^2 \tilde\gamma,K).
\label{eq: cochannel interference capacity proof}
\end{align}
Eq. \eqref{eq: cochannel interference capacity} follows from eq. \eqref{eq: cia simul 1}, \eqref{eq: cochannel interference capacity proof} and Theorem \ref{thm: global MJD}.
\end{proof}
\end{thm}

\subsection{Degrees of freedom}
This section focuses on comparing the degrees of freedom for each of the considered cases. The degrees of freedom determine the number of independent signal dimensions
in the high SNR regime and it is also known as prelog or multiplexing gain in the literature. It is a useful metric in cases where interference is the main impairment and AWGN can be considered unimportant.
\begin{thm}
The degrees of freedom per BS antenna for the global MJD, IA, RDMA and CI cases are given by:
\begin{align}
d_\mathrm{MJD}=1, d_\mathrm{IA}=1-\frac{1}{Mn},d_\mathrm{RD}=1-\frac{1}{2M},d_\mathrm{CI}=1-\frac{1}{M}.
\label{eq: dofs}
\end{align}
\begin{proof}
Eq. \eqref{eq: dofs} can be derived straightforwardly by counting the receive dimensions of the equivalent channel matrices (eq. \eqref{eq: GMJD channel matrix} for global MJD, eq. \eqref{eq: IA channel matrix} for IA, eq. \eqref{eq: RDMA matrix decomposition} and \eqref{eq: RDMA matrix decomposition 2} for RDMA, eq. \eqref{eq: CI channel matrix} for CI) and normalizing by the number of BS antennas. 
\end{proof}
\end{thm} 
\begin{lem}
The following inequalities apply for the dofs of eq. \eqref{eq: dofs}:
\begin{equation}
d_\mathrm{MJD}\geq d_\mathrm{IA}\geq d_\mathrm{RD}>d_\mathrm{CI}.
\end{equation}
\end{lem}
\begin{rem}
It can be observed that $d_\mathrm{IA}=d_\mathrm{RD}$ only for single UT per cell equipped with two antennas ($K=1,n=2$). For all other cases, $ d_\mathrm{IA}> d_\mathrm{RD}$. Furthermore, it is worth noting that when the number of UTs \(K\) and antennas \(n\) grows to infinity, $\lim_{K,n\rightarrow\infty} d_\mathrm{IA}=d_\mathrm{MJD}$ which entails a multiuser gain. However, in practice the number of served UTs is limited by the number of antennas ($n=K+1$) which can be supported at the BS- and more  importantly at UT-side due to size limitations.
\end{rem}
\subsection{Complexity Considerations}
This paragraph discusses the complexity of each scheme in terms of decoding processing and required CSI. In general, the complexity of MJD is exponential with the number of users \cite{Verdu1998} and full CSI is required at the central processor for all users which are to be decoded. This implies that global MJD is highly complex since all system users have to be processed at a single point. On the other hand, clustering approaches reduce the number of jointly-processed users and as a result complexity. Furthermore, CI is the least complex since no action is taken to mitigate intercluster interference. RDMA has an equivalent receiver complexity with CI but in addition it requires coordination between adjacent clusters in terms of splitting the resources. For example, time division would require inter-cluster synchronization, while frequency division could be even static. Finally, IA is the most complex since CSI towards the non-intended BS is also needed at the transmitter in order to align the interference. Subsequently, additional processing is needed at the receiver side to filter out the aligned interference. 
\section{Numerical Results}
\label{sec: numerical results}
This section presents a number of numerical results in order to illustrate the accuracy of the derived analytical expressions for finite dimensions and evaluate the performance of the aforementioned interference mitigation schemes. In the following figures, points represent values calculated through Monte Carlo simulations, while lines refer to curves evaluated based on the analytical expressions of section \ref{sec: eigenvalue distribution analysis}. More specifically, the simulations are performed by generating $10^3$ instances of random Gaussian matrices, each one representing a single fading realization of the system. In addition, the variance profile matrices are constructed deterministically based on the considered \(\alpha\) factors and used to shape the variance of the i.i.d. c.c.s. elements. Subsequently, the per-cell capacities are evaluated by averaging over the system realizations using: a) eq. \eqref{eq: global MJD simul} for global MJD, b) eq. \eqref{eq: first BS signal}-\eqref{eq: IA filtering} and \eqref{eq: IA simul} for IA, c) eq. \eqref{eq: RDMA simul} for RDMA, d) eq. \eqref{eq: cia simul 1},\eqref{eq: cia simul 2} for CI. In parallel, the analytical curves are evaluated based on: a) theorem \ref{thm: global MJD} for global MJD, b) theorems \ref{thm: IA th 1} and \ref{thm: IA th 2} for IA, c) theorems \ref{thm: active RDMA} and \ref{thm: inactive RDMA} for RDMA, d) theorems \ref{thm: global MJD} and \ref{thm: cia thm} for CI.
Table \ref{tab:practicalparas} presents an overview of the parameter values and ranges used for producing the numerical results of the figures. 

\begin{figure}
       \centering
               \includegraphics[width=0.5\textwidth]{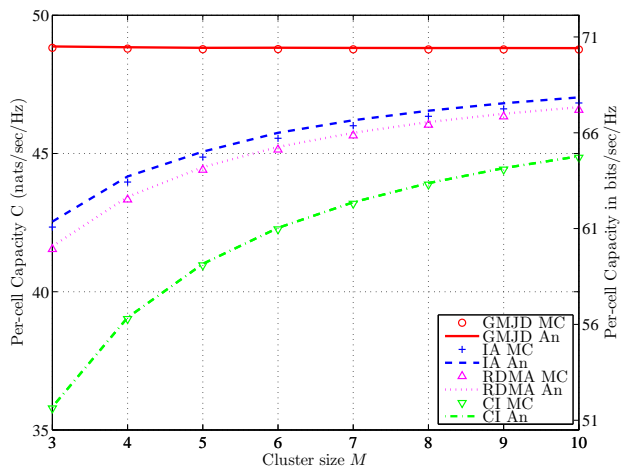}
       \caption{Per-cell throughput scaling versus the cluster size $M$. The performance of global MJD does not depend on the cluster size, while for all interference mitigation techniques, the penalty due to the clustering diminishes as the cluster size increases.}
       \label{fig: capacity vs cluster size}
\end{figure}

\begin{figure}
       \centering
               \includegraphics[width=0.5\textwidth]{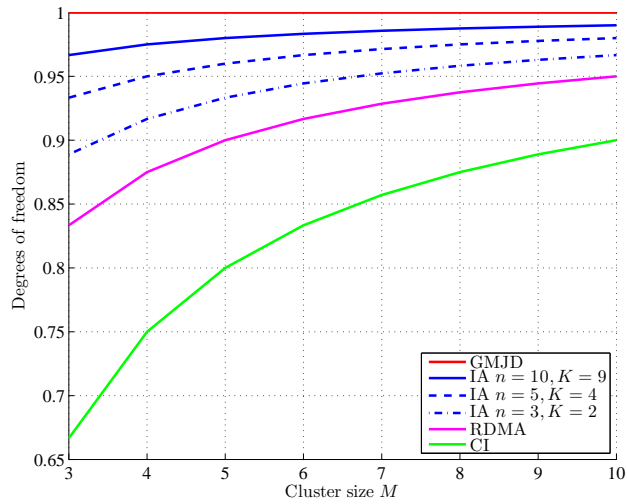}
       \caption{Degrees of freedom versus the cluster size $M$. The IA dofs approach the global MJD dofs as the number of UTs and antennas increases.}
       \label{fig: dof vs cluster size}
\end{figure}

\begin{figure}
       \centering
               \includegraphics[width=0.5\textwidth]{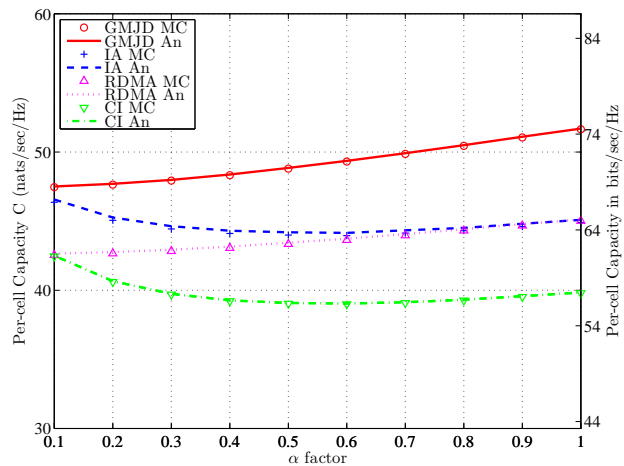}
       \caption{Per-cell throughput scaling versus the $\alpha$ factor. For high $\alpha$ factors, RDMA performance converges to interference alignment, whereas for low $\alpha$ factors RDMA performance degrades even beyond the cochannel interference bound. While the performance of global MJD and RDMA increase monotonically with $\alpha$, the performances of interference alignment and cochannel interference degrade for medium $\alpha$ factors.}
       \label{fig: capacity vs alpha factor}
\end{figure}

\begin{figure}
       \centering
               \includegraphics[width=0.5\textwidth]{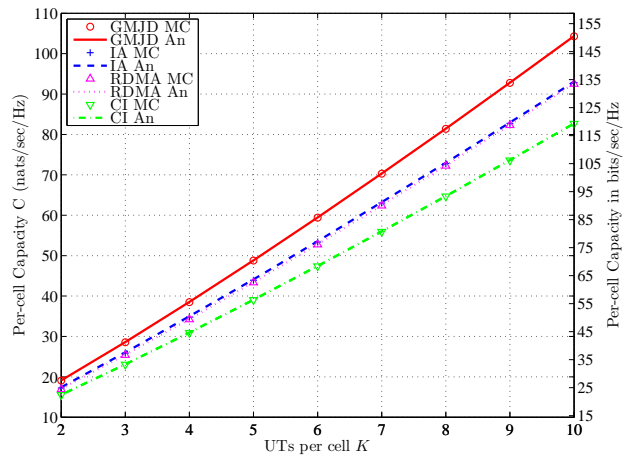}
       \caption{Per-cell throughput scaling vs. the number of UTs per cell $K$. A  superlinear scaling of the performance can be observed, resulting primarily from the increase of spatial dimensions (more antennas) and secondarily from the increase of the system power (more UTs). The slope of the linear scaling is affected by the selected interference mitigation technique.
}
       \label{fig: capacity vs UTs per cell}
\end{figure}

Firstly, figure \ref{fig: capacity vs cluster size} depicts the per-cell throughput versus the cluster size $M$ for medium $\alpha$ factors. It should be noted that the $\alpha$ factor combines the effects of cell size and path loss exponent as explained in \cite{Chatzinotas_Chapter1}. As expected the performance of global MJD does not depend on the cluster size, since it is supposed to be infinite. For all interference mitigation techniques, it can be seen that the penalty due to the clustering diminishes as the cluster size increases. Similar conclusions can be  derived by plotting the degrees of freedom versus the cluster size $M$ (figure \ref{fig: dof vs cluster size}). In addition, it can be observed that the IA dofs approach the global MJD dofs as the number of UTs and antennas increases. Subsequently, Figure \ref{fig: capacity vs alpha factor} depicts the per-cell throughput versus the $\alpha$ factor. For high $\alpha$ factors, RDMA performance converges to IA, whereas for low $\alpha$ factors RDMA performance degrades and touched the CI curve. It should be also noted that while the performance of global MJD and RDMA increase monotonically with $\alpha$, the performances of IA and cochannel interference degrade for medium $\alpha$ factors. Finally, figure \ref{fig: capacity vs UTs per cell} depicts the per-cell throughput versus the number of UTs per cell $K$. It should be noted that the number of antennas per UT $n$ scale jointly with $K$. Based on this observation, a  superlinear scaling of the performance can be observed, resulting primarily from the increase of spatial dimensions (more antennas) and secondarily from the increase of the system power (more UTs). As it can be seen, the slope of the linear scaling is affected by the selected interference mitigation technique.

\section{Conclusion}
\label{sec: conclusion}
In this paper, various techniques for mitigating intercluster interference in clustered MJD were investigated. The case of global MJD was initially considered as an upper bound, serving in evaluating the degradation due to intercluster interference. Subsequently, the IA scheme was analyzed by deriving the asymptotic eigenvalue distribution of the channel covariance matrix using free-probabilistic arguments. In addition, the RDMA scheme was studied as a low complexity method for mitigating intercluster interference. Finally, the CI was considered as a worst-case scenario where no interference mitigation techniques is employed. Based on these investigations it was established, that for dense cellular systems the RDMA scheme should be used as the best compromise between complexity and performance. For average to sparse cellular systems which is the usual regime in macrocell deployments, IA should be employed when the additional complexity and availability of CSI at transmitter side can be afforded. Alternatively, CI could be preferred especially for highly sparse cellular systems. 

\appendices
\section{Proof of Theorem}
\label{app: 1}

\begin{thm}
Let $\mathbf{A}=[\mathbf{0}\ \mathbf{B}\ \mathbf{0}]$ be the concatenation of the variance-profiled Gaussian matrix $\mathbf{B}=\mathbf{C}\odot\mathbf{G}$ and a number of zero columns. Let also $k$  be the ratio of non-zero to total columns of $\mathbf{A}$, $\beta$ be the ratio of horizontal to vertical dimensions of $\mathbf{B}$ and $q$ the Frobenius norm of $\mathbf{C}$ normalized by the matrix dimensions. The R-transform of $\mathbf{A}^H\mathbf{A}$ is given by:
\begin{align}
\mathcal{R}_{\frac{1}{n}\mathbf{A}^H\mathbf{A}}(z,k,\beta,q)&={\frac {k-zqk(\beta +1)\pm\sqrt {{k}^{2} \left( {q}^{2}{\beta}^{2}{z}^{2}-2
\,q\beta z-2\,{z}^{2}{q}^{2}\beta+1-2\,qz+{z}^{2}{q}^{2}+4\,zqk \right)}}{2z\left( q\beta z-k \right) }}
\end{align}
\label{thm: concatenated R transform}
\begin{proof}
%Variance-profiled Gaussian matrices concatenated with zero columns. Shannon transform is unaffected by zero eigenvalues since $\mathcal{V}(0)=0$. Stieltjes and R- transforms are affected.
%
Let $\mathbf{B}=\mathbf{C}\odot\mathbf{G}_{n\times m}$ be a variance-profiled Gaussian matrix with $\beta={m}/{n}$ and $q=\Vert\mathbf{C}\Vert^2/nm$. According to \cite{Chatzinotas_JWCOM}, the R-transform of ${\frac{1}{n}\mathbf{B}^H\mathbf{B}}$ is given by:
\begin{equation}
\mathcal{R}_{\frac{1}{n}\mathbf{B}^H\mathbf{B}}(z)=\frac{q}{1-\beta q z}.
\label{eq: variance R-transform}
\end{equation}
Using eq. \eqref{eq: Cauchy}, the Stieltjes transform of ${\frac{1}{n}\mathbf{B}^H\mathbf{B}}$ can be expressed as:
\begin{equation}
\mathcal{S}_{\frac{1}{n}\mathbf{B}^H\mathbf{B}}(z)={\frac {-z+q-q\beta\pm \sqrt {{z}^{2}-2\,zq-
2\,zq\beta+{q}^{2}-2\,{q}^{2}\beta+{q}^{2}{\beta}^{2}} }{2zq\beta}}.
\end{equation}
Matrix ${\frac{1}{n}\mathbf{A}^H\mathbf{A}}$ has identical eigenvalues to ${\frac{1}{n}\mathbf{B}^H\mathbf{B}}$ plus a number of zero eigevalues with $0<k<1$ defined as the ratio of non-zero eigenvalues over the total number of eigenvalues. As a result, the a.e.p.d.f. of ${\frac{1}{n}\mathbf{A}^H\mathbf{A}}$ can be written as:
\begin{equation}
f_{\frac{1}{n}\mathbf{A}^H\mathbf{A}}(z)=k f_{\frac{1}{n}\mathbf{B}^H\mathbf{B}}(z)+(1-k)\delta(x).
\end{equation}
Using the definition of the Stieltjes transform \cite{Tulino04}:
\begin{equation}
\mathcal{S}_{\frac{1}{n}\mathbf{A}^H\mathbf{A}}(z)=k \mathcal{S}_{\frac{1}{n}\mathbf{B}^H\mathbf{B}}(z)-\frac{1-k}{z}
\end{equation}
and employing eq. \eqref{eq: Cauchy}, the proof is complete.
\end{proof}
\end{thm}


%\section*{Acknowledgment}

\ifCLASSOPTIONcaptionsoff
  \newpage
\fi



\bibliographystyle{IEEEtran}
\bibliography{IEEEabrv,references,journals,books,conferences,thesis}

\newpage

\begin{table}
\caption{Parameters for throughput results}
\centering
\begin{tabular}{l|c|c|c|r}
Parameter & Symbol & Value & Range & Figure\\\hline
Cluster size  & \(M\) & $4$ & $3-10$ & \ref{fig: capacity vs cluster size}\\
$\alpha$ factor & \(\alpha\) & $0.5$ & $0.1-1$ & \ref{fig: capacity vs alpha factor}\\
UTs per cell  & \(K\) & $5$ & $2-10$ & \ref{fig: capacity vs UTs per cell}\\
Antennas per UT  & \(n\) & $4$ & $3-11$ & \ref{fig: capacity vs UTs per cell}\\
UT Transmit Power & \(\gamma\) & $20dB$ & &\\
Number of MC iterations &   & $10^3$ & &\\\hline
\end{tabular}
\label{tab:practicalparas}
\end{table}

\end{document}